\documentclass[fleqn,10pt]{wlscirep}
\usepackage[utf8]{inputenc}
\usepackage[T1]{fontenc}

\title{Optimization of DMD-based \textcolor{black}{independent amplitude and phase} modulation: a spatial resolution and quantization}

\usepackage{amsmath,amsfonts,amssymb}

\author[1,*]{Alexandra Georgieva}
\author[1,2]{Andrey Belashov}
\author[1]{Nikolay V. Petrov}
\affil[1]{ITMO University, Department of Photonics and Optical Information Technologies, Saint-Petersburg, 199004, Russia}
\affil[2]{Ioffe Institute, Saint-Petersburg, 194021, Russia}

\affil[*]{georgieva@itmo.ru}

\begin{abstract}
This paper presents the results of a comprehensive study on the optimization of the optical system for independent amplitude and phase wavefront manipulation, which is based on a digital micromirror device implementation. The parameters of generated binary fringe patterns to reach the optimal quality of target complex-valued wavefront have been found for several examples, and a general algorithm for DMD pattern optimization is described. It was shown that a trade-off between spatial resolution and quantization of the target amplitude and phase distribution should be achieved. The increase of carrier frequency results in higher spatial resolution of the modulated complex wave but decreases its quantization. The dependence of the best DMD-pattern generation parameters on the type of target complex wavefront is discussed in terms of spatial resolution and quantization degree. 
\end{abstract}

\begin{document}

\flushbottom
\maketitle

\thispagestyle{empty}

\section*{\fontsize{10}{15}\selectfont \textit{Keywords}}
Digital micromirror device; Wavefront modulation; Digital holography; Spatial resolution; Image quantization

\section{Introduction}

The synthesis of wavefronts with known characteristics has gained the interest of many researchers in the field of photonics. Some of the \textcolor{black}{tasks} %applications
of wavefront \textcolor{black}{shaping} %modulation 
are high-resolution microscopy~\cite{booth2015aberrations}, laser beam shaping~\cite{dickey2018laser,ouadghiri2016arbitrary}, scattering media characterization~\cite{yu2017ultrahigh,horstmeyer2015guidestar,coyot2015implementation}, holographic displays~\cite{chipala2019color}, quantum cryptography~\cite{mirhosseini2015high}, metrology~\cite{aulbach2017NonContact}, compressed sensing~\cite{gao2014single}, 3D bioprinting and lithography~\cite{yoon2018emerging}. To date, there exists a range of static and dynamic wavefront modulators, such as diffraction optical elements~\cite{rohwetter2008laser}, metasurfaces~\cite{wang2016subwavelength}, adaptive optical elements ~\cite{salter2019adaptive}, which provide the possibility to operate with the amplitude, phase, or polarization of the beam profile in a wide range of wavelengths~\cite{wu2020DMD,he2020metasurfaces}. Adaptive spatial light modulators with programmable precise control of the wavefront have become a valuable tool for various applications, e.g., in imaging systems~\cite{zhou2020single}. Two types of such devices can be outlined: liquid crystal-based spatial light modulators and micro-electromechanical systems (MEMS). The former includes such subtypes as transmissive liquid crystal, reflective liquid crystal on silicon, and ferroelectric liquid crystal. MEMS-based spatial light modulators are presented by digital micromirror device (DMD), active micromirror matrix, and grating light valve~\cite{zlokazov2020methods}.

Each of the \textcolor{black}{devices characterized by the type of modulation, among which they are distinguished: purely amplitude, pure-phase, and simultaneous amplitude-phase modulation, and each of} modulators has its benefits and disadvantages~\cite{yu2017wavefront,turtaev2017comparison,zlokazov2020methods}. The choice of the required device is determined by \textcolor{black}{the peculiarities of the problem to be solved} %the parameters necessary
in a particular case. Several important characteristics of wavefront modulators can be highlighted: the speed of operation, dynamic range, number and size of pixels, modulation efficiency. In applications where high speed is required and spatial resolution can be sacrificed to achieve high light modulation rates~\cite{ren2015tailoring}, the use of DMD, constructively assuming only binary modulation, is preferable due to its high refresh rate (22 kHz). Over the past few years, such devices have been actively used in various studies~\cite{ren2015tailoring,yoon2018emerging}, and commercial devices (e.g., holotomographic microscope, developed by Tomocube\texttrademark). Compared to other modulators, DMD has high switching speed%(microseconds)
, high light efficiency, high fill-factor (90\%), and relatively low cost~\cite{park2015properties,kalyoncu2013fast,feng2017digital}. It provides a high enhancement factor in the task of focusing through the scattering medium or improving contrast in optical imaging and high beam-shaping fidelity~\cite{yu2017wavefront,turtaev2017comparison}. This is particularly relevant in biomedical applications where rapid processes are involved, modulation of ultrashort radiation, or the possibility of real-time measurement should be provided~\cite{ren2015tailoring,yoon2018emerging,wen2019spatially}. DMD consists of a CMOS-placed micromirrors array, each of which can have only two stable states: ``\textcolor{black}{O}n'' ($+12^{\circ}$) and ``\textcolor{black}{O}ff'' ($-12^{\circ}$)~\cite{ren2015tailoring}. Each micromirror represents a single pixel of the projected image. In addition, the use of binary (1-bit) holograms is convenient in terms of data capacity, for example, for implementation in holographic displays~\cite{pan2016review}. Another advantage of binary holograms over grayscale holograms is also that they can be easily printed~\cite{tsang2011computer}.

Various approaches have previously been proposed for generating binary \textcolor{black}{DMD-patterns} or converting grayscale holograms into binary holograms, for instance, global and local thresholding methods~\cite{cheremkhin2019comparative}, iterative techniques~\cite{chhetri2001iterative,stuart2014fast}, error diffusion method~\cite{yang2019error}, the superpixel-based method~\cite{goorden2014superpixel}, off-axis computer-generated Lee holography technique~\cite{lee1974binary}. The latter is an efficient and fast method, especially eligible for ultrafast radiation modulation~\cite{gu2015digital}.
In this method, the $1^{st}$ diffraction order Fourier filtering is utilized, commonly with an aperture in experimental settings, which affects spatial resolution. The number of available amplitude levels (or image quantization) depends on the number of pixels in the fringe period in DMD-generated pattern~\cite{conkey2012high}. These parameters significantly affect the image quality. Various methods of quantization and resolution improvement were proposed. The study by Reimers et.al.~\cite{reimers2017freeform} has elucidated the optimal resolution obtaining for different object detection in terms of hyperspectral imaging. The results reported by Zhang et. al.~\cite{zhang2017fast} suggest that in single-pixel imaging, the quantization error caused by binarization can be eliminated by error diffusion dithering and a high ratio of upsampling. In their \textcolor{black}{recent} study, Chipala and Kozacki~\cite{chipala2019color} demonstrated the correlation between the dispersion of the DMD and the resolution of holographic images, and proposed the method of image quality improvement. However, the area of achieving maximum quantization and resolution values by optimizing the experimental setup and binarization parameters has not been explored in depth. 

Thus, \textcolor{black}{in a given work, we developed} an approach \textcolor{black}{for} the optimization of \textcolor{black}{the technique of independent amplitude and phase modulation based on DMD} and binarization parameters to obtain the optimal spatial resolution and quantization depending on the target complex wave. Binary DMD patterns, generated by the Lee method~\cite{lee1974binary}, and the experimental setup of off-axis holography were used to investigate the spatial resolution and image quantization depending on the number of pixels per fringe period and aperture size for the $1^{st}$ diffraction order filtration. To obtain the optimal parameters, the root mean square error (RMSE) for each case was calculated. In addition, experimental validation of the proposed optimization method was performed. It was established that for applications requiring high spatial resolution, it was necessary to sacrifice the number of amplitude \textcolor{black}{gradations} to increase the filtration aperture in Fourier domain. The work also clearly demonstrates that for applications where it is important to preserve the gradation, it is possible to increase the binary fringe period of DMD-pattern.

The paper is organized as follows: In Section~\ref{sec:methods}, an experimental setup and principles of \textcolor{black}{independent amplitude and phase} wavefront modulation using \textcolor{black}{binary} DMD are described. In Section~\ref{sec:results}, the results of a detailed study of the dependence of the modulated wavefront quality on the parameters of wavefront modulation are presented. The spatial resolution and image quantization is considered depending on the filtration aperture size and carrier frequency in Subsection~\ref{subsec:filtration}. In Subsection~\ref{subsec:RMSE}, the calculation of the optimal aperture and binary fringe period is demonstrated, and Subsection~\ref{subsec:experiment} shows experimental results of wavefront manipulation using random and optimized DMD pattern.

\section{Methods}
\label{sec:methods}
\subsection{Experimental setup}
\label{subsec:setup}
\begin{figure} [ht]
   \begin{center}
   \begin{tabular}{c} 
   \includegraphics[height=7cm]{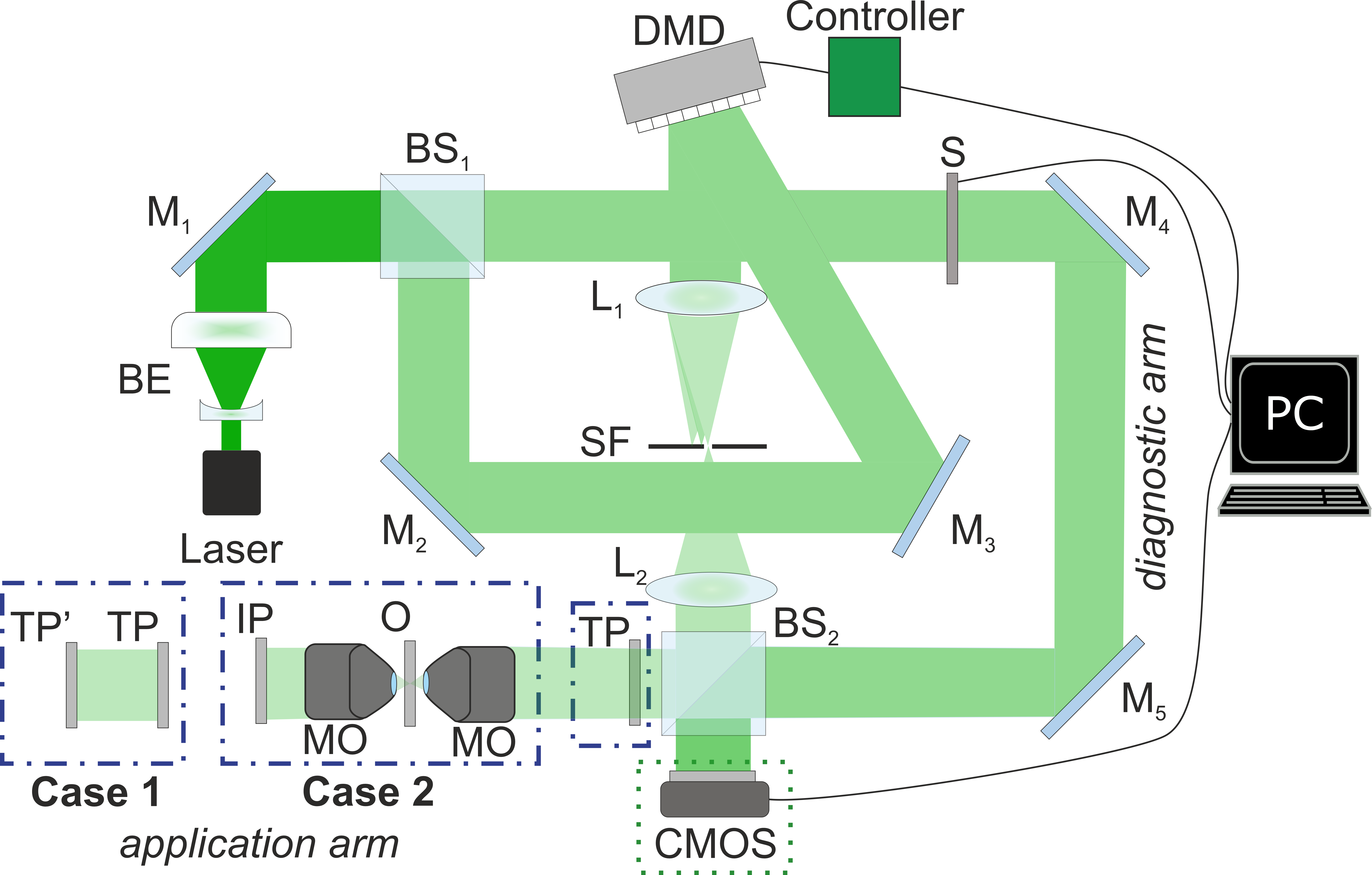}
   \end{tabular}
   \end{center}
   \caption{Experimental setup. \textsf{BE} are beam expanders, \textsf{M}$_{1-5}$ are the mirrors, \textsf{S} is a shutter, \textsf{BS}$_{1-2}$ are the beam splitters, \textsf{L}$_{1-2}$ are lens, \textsf{SF} is the spatial filter, \textsf{TP} is the target plane, \textsf{CMOS} is the matrix detector, MO are objectives, IP is an image plane, O is an object.} 

    \label{fig:setup} 
   \end{figure}

The experimental setup is represented by the Mach-Zender interferometer with DMD in the object beam (Fig.~\ref{fig:setup}).

This setup provides an opportunity to modulate the wavefront using DMD, as well as to evaluate its modulation quality by image reconstruction with off-axis digital holograms. DMD was operated using controller DLPC900 (modulation frequency 9 kHz (1-bit image), 1 kHz (8-bit image)). The laser beam with the wavelength of 532~nm was \textcolor{black}{increased in diameter} by the beam expander~\textsf{BE}, then it was divided into object and reference waves by the beam splitter \textsf{BS}$_1$. 
The object beam \textcolor{black}{was} incident on the DMD \textcolor{black}{(Texas Instrument Light Crafter DLP6500FYE in our case with 1920$\times$1080 micromirrors with a size of $7.56~\mu$m)}, where the corresponding pattern \textcolor{black}{was displayed on the  matrix.} Then it was focused with the lens \textsf{L}$_1$ to the Fourier plane, where the 1$^{st}$ diffraction order \textcolor{black}{was} separated by the spatial filter \textsf{SF} with aperture size scaleable along the $x$ coordinate. Afterward, the lens \textsf{L}$_2$\textcolor{black}{, which together with \textsf{L}$_1$ forms telecentric 4$f$-system}, collimated the radiation. The desired field distribution was formed in \textcolor{black}{the target plane}~\textsf{TP} \textcolor{black}{after the second beamsplitter \textsf{BS}$_2$. The presence of \textsf{BS}$_2$ ensured the} location of this plane simultaneously both at the end of the diagnostic arm, namely, in the sensor plane (\textsf{CMOS}) and in the application arm.  
\textcolor{black}{The beamsplitter \textsf{BS}$_2$ enabled \textcolor{black}{performing both the} simultaneous monitoring of the wave, \textcolor{black}{independently modulated in amplitude and phase} by off-axis digital holography using \textsf{CMOS} sensor in the monitoring arm (indicated with \textcolor{black}{green dotted frame} in Fig.~\ref{fig:setup}) and utilized the modulated wave for any research purposes in \textcolor{black}{the} application arm (indicated with \textcolor{black}{blue dash-dotted frame} in Fig.~\ref{fig:setup}. Plane reference wave (when the shutter \textsf{S} is opened) passes through the mirrors \textcolor{black}{\textsf{M}$_{4}$ and \textsf{M}$_{5}$} and enabled detection of \textcolor{black}{formed} amplitude and phase distributions in the application arm.} In our study, we used the local least square estimation algorithm\cite{liebling2004complex,katkovnik2015wavefront} for wavefront reconstruction. 

\textcolor{black}{It should be noted that some tasks may require to have the wavefront of a specific structure beyond the target plane. This wavefront may be engineered very precisely by means of an analysis of the resulting field at~\textsf{TP} in the diagnostic arm and solving the diffractive equations, which describe the wavefront propagation beyond the~\textsf{TP}~\cite{Dudley:14} (case 1 in  Fig.~\ref{fig:setup}). Then, this information should be taken into account in the iterative feedback loop. The alternative is not to propagate wavefront numerically, but to image \textsf{TP} further by an additional telecentric system with custom magnification~\cite{ding2020wavefront} (case 2 in Fig.~\ref{fig:setup}). It is also necessary to mention that independent values in amplitude and phase distributions may be obtained in the \textsf{TP} \textcolor{black}{(or in the corresponding image plane, as a result of the projection of the 4-$f$ telecentric system)} only, since the diffraction leads to the projection of phase structure on the amplitude.}

\subsection{Principles of binary complex wavefront modulation}
\label{subsec:principles}

In this subsection, the basic principles of independent amplitude and phase modulation of the wavefront using DMD generated binary off-axis digital hologram are considered and discussed. One of the possible approaches for independent amplitude and phase modulation was proposed by Lee et al.~\cite{lee1974binary} is to generate a binary pattern and ``reconstruction'' of the synthetic off-axis digital hologram.

It is known that fringes in an analog hologram can be used to change the phase and amplitude of the incident light wave, thus producing an image of the recorded object.
Similar to analog holography, the variation of binary fringe parameters such as width and periodicity in the formed binary DMD pattern, enables manipulation of amplitude and phase distributions.
The approach to obtaining independent amplitude-phase modulation from binary amplitude modulation is based on the spatial filtering of the wave in one of the diffraction orders arising from the reflection of an incident wave from DMD.
\textcolor{black}{It can be conducted, for example, by spatial filtration of this diffraction order if the binary pattern is an off-axis hologram with high enough carrier frequency.}

Experimentally, it is usually implemented by Fourier transform performed using a concave lens (\textsf{L1}), as demonstrated in Fig.~\ref{fig:setup}. Filtration of the first diffraction order using adjustable iris aperture (\textsf{SF}), while inverse Fourier transform by another concave lens L2 enables the reconstruction of the target wave modulated by DMD.

The spatial frequency of the binary hologram is calculated using the carrier frequency and phase change as follows~\cite{lee1974binary}:
\begin{equation}
    \label{eq:sp_freq}
\begin{aligned}
\nu_x(x,y)=k_x+\frac{1}{2\pi}\frac{\partial \phi(x,y)}{\partial x}\\
\nu_y(x,y)=k_y+\frac{1}{2\pi}\frac{\partial \phi(x,y)}{\partial y},
\end{aligned}
\end{equation}
where $k_x$ and $k_x$ are carrier frequencies in the $x$ and $y$ directions, which are inversely proportional to the specified fringe period due to the wavefront tilt. The second summand refers to the spatial frequency of the wave with target phase distribution. 

\begin{figure}[ht]
\begin{center}
   \begin{tabular}{c} %% tabular useful for creating an array of images 
   \includegraphics[height=10.5cm]{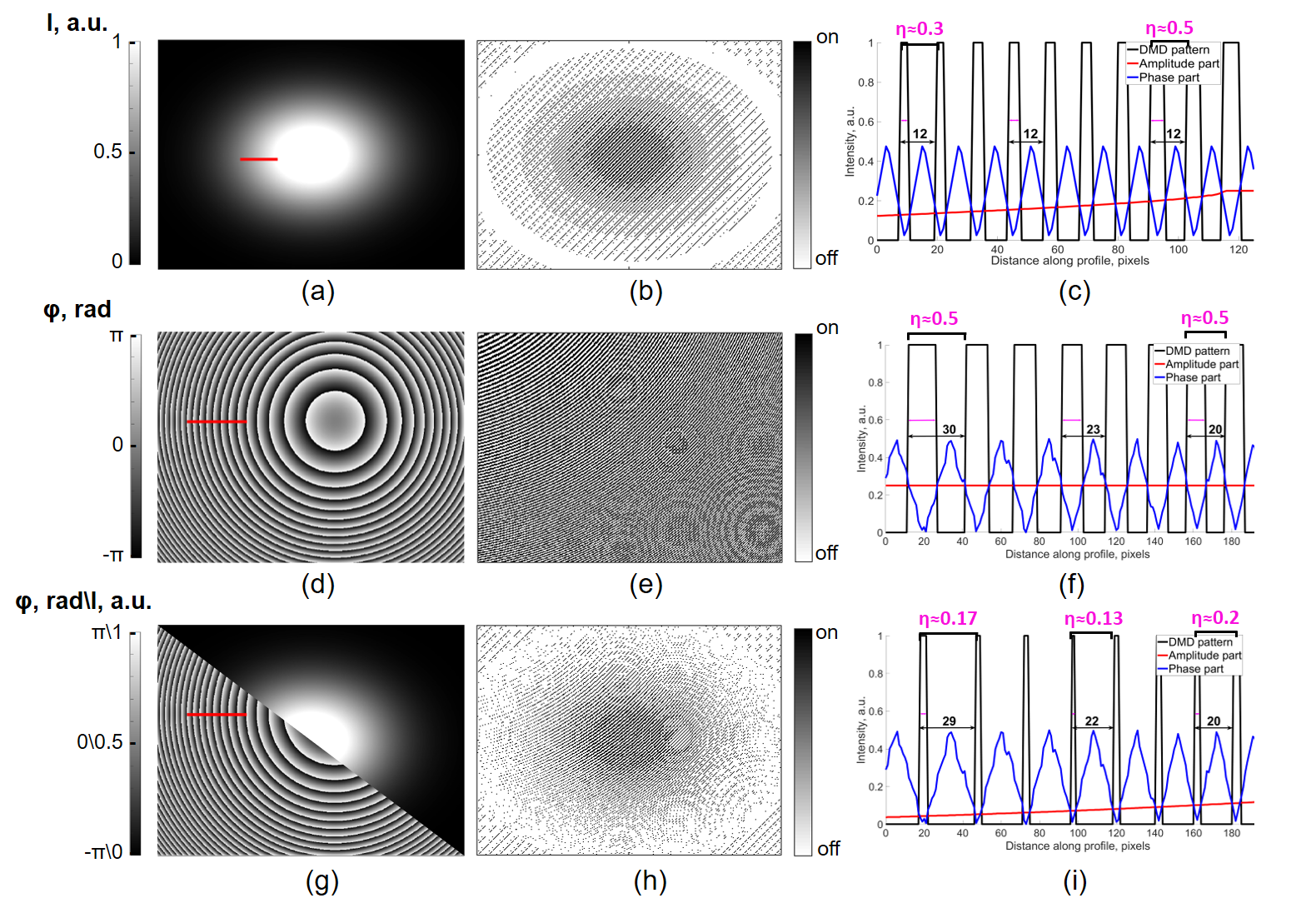}
   \end{tabular}
   \end{center}
    \caption{Demonstration of independent amplitude (a-c), phase (d-f), and complex (g-i) modulation. Target amplitude (a), phase (d), complex (g) distribution. The red line indicates the coordinates of plotting in (c,$\,$f,$\,$i); (b,$\,$e,$\,$h) Resulting DMD pattern (carrier frequency $\frac{2\pi}{12}$); (c,$\,$f,$\,$i) the plot of DMD pattern function (black line), phase part (blue line) and amplitude part (red line).}
    \label{fig:principles}
\end{figure}

DMD-pattern $h(x,y)$ was formed using the limiter to obtain amplitude hologram binarization~\cite{lee1974binary}:
\begin{equation}
\label{eq:ineq}
h(x,y)=
    \begin{cases}
    1; \quad  |([\frac{\phi(x,y)}{2\pi}+(k_x \cdot x + k_y \cdot y)] \mod 1) -0.5|\le\frac{asin|A(x,y)|}{2\pi}
    \\
    0;   \quad \text{otherwise}
    \end{cases}
\end{equation}
where $A(x,y)$ is the target amplitude, $\phi(x,y)$ is the target phase, $k_x$ and $k_y$ are the carrier frequencies in DMD pixels of the binary off-axis hologram in the $x$ and $y$ direction. The left part of the inequality~\eqref{eq:ineq} is responsible for the encoding of the phase distribution, while the amplitude distribution is encoded by the right part. To show this segregation of amplitude and phase parts, we consider DMD-pattern simulated for amplitude-only, phase-only, and complex-valued (simultaneous independent amplitude and phase) modulations (Fig.~\ref{fig:principles}).

The first example of varying amplitude (2D Gaussian distribution) and constant phase distributions is presented in Fig.~\ref{fig:principles}~(a-c). The target amplitude image (Fig.~\ref{fig:principles}~(a) and the simulated binary pattern (Fig.~\ref{fig:principles}~(b)) demonstrate how the complex wave amplitude can be modulated by the variation of the relative amount of DMD pixels at On-state (the On-pixels). The plot of phase part of inequality, indicated by the red line, clearly shows how the inequality~\eqref{eq:ineq} provides an opportunity to readily synthesize a binary pattern for a given amplitude distribution. The right (amplitude) and left (phase) parts of the inequality are presented by red and blue lines, respectively. While in this case, the phase distribution is constant, the left part of the inequality representing the phase distribution, is a simple periodical function due to the linear increase of $\frac{\phi(x,y)}{2\pi}+(k_x \cdot x + k_y \cdot y)$ term and '$mod 1$' operation. However, the target amplitude image is changed with $x$ coordinate (red line), thus varying the relative amount of the On and Off pixels. If the amplitude part is larger than the phase part, the pixel is On, otherwise it remains Off. In this case, the ratio between the On pixels and the total number of pixels per one period (the On pixel occupancy coefficient $\eta$) is changed from 0.3 to 0.5 while the target amplitude increases from $\approx $0.13 to $\approx $0.25.

Due to the constant phase distribution, the carrier frequency, as well as the value of spatial frequency of binary hologram here is constant and equals $\frac{2\pi}{12}$. Subsequently, the fringe period is 12 DMD pixels (indicated with black arrows in Fig.~\ref{fig:principles}~(c)).
%%%%%%%%%%%%%%%%%%%%%%%%%%%%% Principles image
%%%%%%%%%%%%%%%%%%%%%%%%%%%%% Principles image
The phase-only variation is demonstrated with the target spherical phase distribution and constant amplitude. Fig.~\ref{fig:principles}~(d-f) show the target phase distribution (Fig.~\ref{fig:principles}~(d)), the simulated DMD pattern (Fig.~\ref{fig:principles}~(e)) and the plotted functions of DMD pattern pixels, amplitude part and phase part (Fig.~\ref{fig:principles}~(f)) of inequality~\eqref{eq:ineq} in the case of phase-only modulation. In this example, the right part of the inequality~\eqref{eq:ineq} is constant due to the absence of amplitude modulation. However, variation of the target phase along $x$ axis results in the change of $\frac{\phi(x,y)}{2\pi}+(k_x \cdot x + k_y \cdot y)$ slope and alteration of the phase part periodicity and spatial frequency of binary hologram.
The black arrows in Fig.~\ref{fig:principles}~(f) demonstrate how the phase variation results in the change of the fringe period from 29 to 20 pixels. At the same time, $\eta$ is constant among the binary hologram and equal to 0.5.

Therefore, phase modulation is achieved by the variation of the binary fringe period, while amplitude modulation is obtained due to the variation of the On pixel occupancy coefficient. Since it is possible to separately vary these two parameters in each local area of the image, the independent manipulation of the phase and amplitude distributions can be achieved.

An example of independent modulation of both amplitude and phase parts of the complex wavefront is presented in Fig.~\ref{fig:principles}~(g-i). It can be noted that both fringe periodicity and the On pixel occupancy coefficient are changed due to amplitude and phase variation.

\section{Results}
\label{sec:results}
\subsection{Experimental settings impact on the quality of the resulting wavefront}
\label{subsec:filtration}

\textcolor{black}{It was previously described how a binary pattern is generated for independent modulation of amplitude and phase distributions. However, for practical applications, the quality of the resulting wavefront, which is influenced by the experimental setup parameters, is crucial. One of the most important steps of modulation is the spatial filtration of one diffraction order. A concave lens is usually utilized for this purpose, in the focal plane of which the filtering is performed. This is the most common approach, which was used, e.g., in~\cite{ren2015tailoring,shin2016optical}.}

In this section, the modulated wavefront quality is discussed and analyzed as a function of the $1^{st}$ order filtration aperture size and carrier frequency of binary hologram.
To demonstrate the impact of experimental parameters on quality of the reconstructed wavefront, the ``mountain'' and ``sky'' images were taken as target amplitude and phase distributions. These two images were chosen since the impact of both gray levels quantization and spatial resolution can be visible on them %on sky and mountain images 
 as they have both smoothly and sharply varying gray levels objects.

\begin{figure}[ht]
   \begin{center}
   \begin{tabular}{c}  
   \includegraphics[height=9cm]{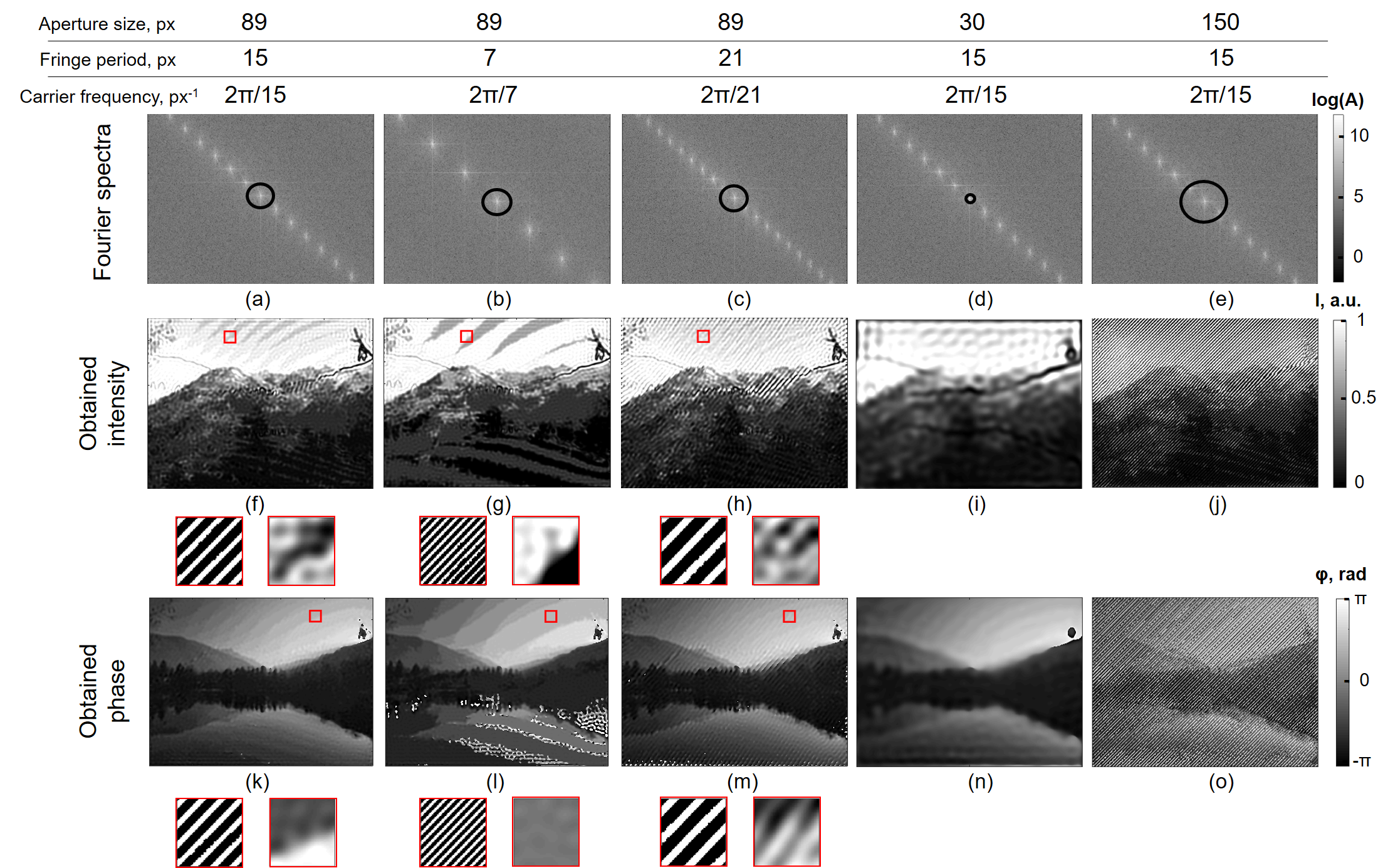}
   \end{tabular}
   \end{center}

\caption{Dependence of reconstructed image quality on aperture size and fringe period (carrier frequency). Fourier spectra with indicated 1$^{st}$ order filtration apertures of different sizes (black circles) (a\,--\,e). The spectra are shifted so that the 1$^{st}$ diffraction order is located in the center; obtained intensity distributions (f\,--\,j). Obtained phase distributions (k\,--\,o). Insets: enlarged fragments of corresponding DMD patterns and fragments of phase or intensity distributions.}
   \label{fig:ap_per_change}
  \end{figure}

If the target amplitude and phase distributions have a large number of small details, the 1$^{st}$ order filtration can be a challenging problem due to the presence of high spatial frequencies overlapping with other diffraction orders. In this case, target wavefront blurring will be observed due to the inevitable loss of high spatial frequencies and corresponding data of small details. However, even under these conditions, it is possible to find the optimal aperture size, which enables obtaining a wavefront with minimal error. Fig.~\ref{fig:ap_per_change} demonstrates the impact of filtration aperture size and carrier frequency or fringe period variation on the generated complex wave and reveals the relationship between these parameters and image quality.

The first column of Fig.~\ref{fig:ap_per_change} represents the case with an optimal aperture size and fringe period. Very small filtration area (Fig.~\ref{fig:ap_per_change}(d)) eliminates high spatial frequencies, which results in image blurring, as it can be seen in the fourth column (Fig.~\ref{fig:ap_per_change}~(i,$\,$n)). On the other hand, excessively large aperture size (Fig.~\ref{fig:ap_per_change}~(e)) leads to the adjacent diffraction orders penetration through the filtration aperture and the image quality reduction by defect appearance (Fig.~\ref{fig:ap_per_change}~(j,$\,$o)).

To simplify the filtration and increase the aperture size, the distance between diffraction orders can be increased by the increment of the carrier frequency. The dependence between binary fringe periodicity and image quality was analyzed during carrier frequency variation. The second and third columns of Fig.~\ref{fig:ap_per_change} demonstrate filtration of the $1^{st}$ diffraction order in Fourier plane and obtained amplitude and phase distributions at various values of carrier frequency. These results show a significant improvement in spatial resolution due to the carrier frequency increase. However, the increase of carrier frequency is associated with the decrease of the average fringe period as well. As was demonstrated above and shown in Fig.~\ref{fig:principles}, modulation of the wavefront amplitude and phase is possible due to the local variation of (1) the On pixel occupancy coefficient and (2) fringe period. The latter is related with the \textcolor{black}{image quantization} and its alteration results in the inevitable variation of the number of amplitude levels of the obtained distribution, as it can be seen in Fig.~\ref{fig:ap_per_change}~(g,$\,$l,$\,$h,$\,$m). 
The explicit demonstration of the correlation between the fringe period and the image quantization is shown in insets for intensity distributions. 
Since the amplitude modulation is based on the variation of the On pixel occupancy coefficient, only six combinations of amplitude values can be encoded for fringe period of 7, while for the fringe period of 21 enables the encoding of 20 various values of the target amplitude (see insets for Fig.~\ref{fig:ap_per_change}~(f,$\,$g,$\,$h)). Similarly, the phase value in the local area can be changed by variation of the fringe period. However, if the typical fringe period is small, it cannot be neither significantly decreased nor increased, which results in limited capability for phase value variation and low quantization of phase shift. Therefore, in the case of lower carrier frequency (larger fringe period), the image quantization level is high. 
This effect is also demonstrated in insets for the obtained phase distributions. For example, in insets for Fig.~\ref{fig:ap_per_change}~(l) only one phase value is encoded (one color), while for insets for Fig.~\ref{fig:ap_per_change}~(k), it is clear that there are at least two phase values. Furthermore, the fringe period in inset for Fig.~\ref{fig:ap_per_change}~(m) changes by 1-2 pixels, allowing the smaller phase variations to be encoded. Therefore, increasing the fringe period (reducing carrier frequency) enables encoding smaller variations in the amplitude and phase of the wavefront, while complicating the first diffraction order filtering.

Thus, it was shown that the increase of carrier frequency results in higher spatial resolution but lower amplitude and phase quantization. Therefore, in the general case of DMD-based wavefront manipulation, the trade-off between two image quality compounds, namely, spatial resolution and quantization, should be sustained. If high spatial resolution is required for the modulated wavefront, a large enough carrier frequency of the binary pattern should be chosen to increase the distance between diffraction orders. However, if the target complex wave suggests smoothly varying amplitude and phase distributions, the carrier frequency should be limited to increase amplitude and phase quantization. In the next subsection, several examples of various amplitude and phase images encoded in a binary pattern will be presented.
Moreover, such a trade-off between these two image properties can be easily understood as the binary diffraction pattern may contain a limited amount of initial information about the target wavefront. More information can be provided either about spatial resolution (by the increase of carrier frequency) or quantization (by the decrease of carrier frequency), although both of these complex wave characteristics cannot be improved at the same time.

\subsection{Optimization of aperture size and carrier frequency for modulation error reduction}
\label{subsec:RMSE}

The results presented above demonstrate a strong correlation between image quality and aperture size and carrier frequency or fringe period. In dependence on the target wavefront parameters, various binary patterns with different carrier frequencies should be generated. In this section, we consider the optimization of wavefront modulation quality for several representative examples of target complex waves. To perform such optimization, we numerically simulated DMD-based target wavefront modulation with various carrier frequencies and filtration aperture sizes and calculated the wavefront modulation error as root mean square error (RMSE) from the target amplitude or phase distributions. Minimization of the calculated error enables obtaining the optimal parameters of DMD-generated binary pattern for a particular target complex wave.

As it will be shown later, the encoding of sharp image elements such as resolution test charts requires high spatial resolution, which can be achieved by using a large spatial aperture size. In this case, a small number of amplitude or phase levels is usually required for successive modulation of the target complex wave. On the other hand, in the case of relatively smooth objects, amplitude or phase quantization is a more important parameter than spatial resolution. Thus, the carrier frequency as well as aperture size can be decreased in this case. 

\begin{figure}[h!]
  \begin{center}
   \begin{tabular}{c} %% tabular useful for creating an array of images 
   \includegraphics[height=8.5cm]{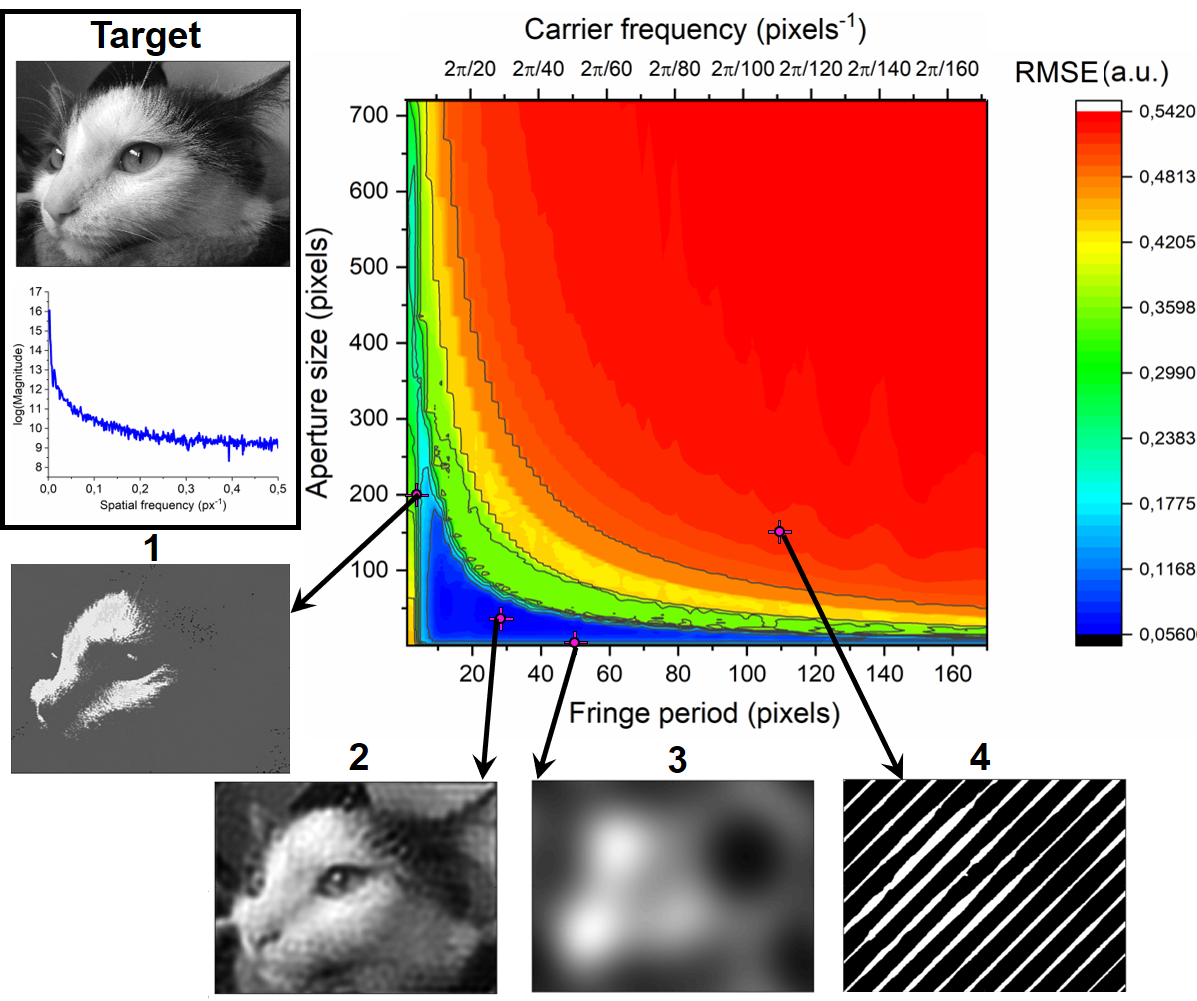}
   \end{tabular}
   \end{center} 
\caption{Amplitude modulation RMSE map dependent from aperture size and fringe period for the ``cat'' object. Insets: obtained intensity distributions at the points indicated by the markers. \textcolor{black}{Target intensity distribution with corresponding angular spectrum structure diagram are shown in a black rectangle.}}
%>>>> use \label inside caption to get Fig. number with \ref{}
   \label{fig:cat_error}
  \end{figure}

\begin{figure}[h!]
  \begin{center}
   \begin{tabular}{c} %% tabular useful for creating an array of images 
   \includegraphics[height=9cm]{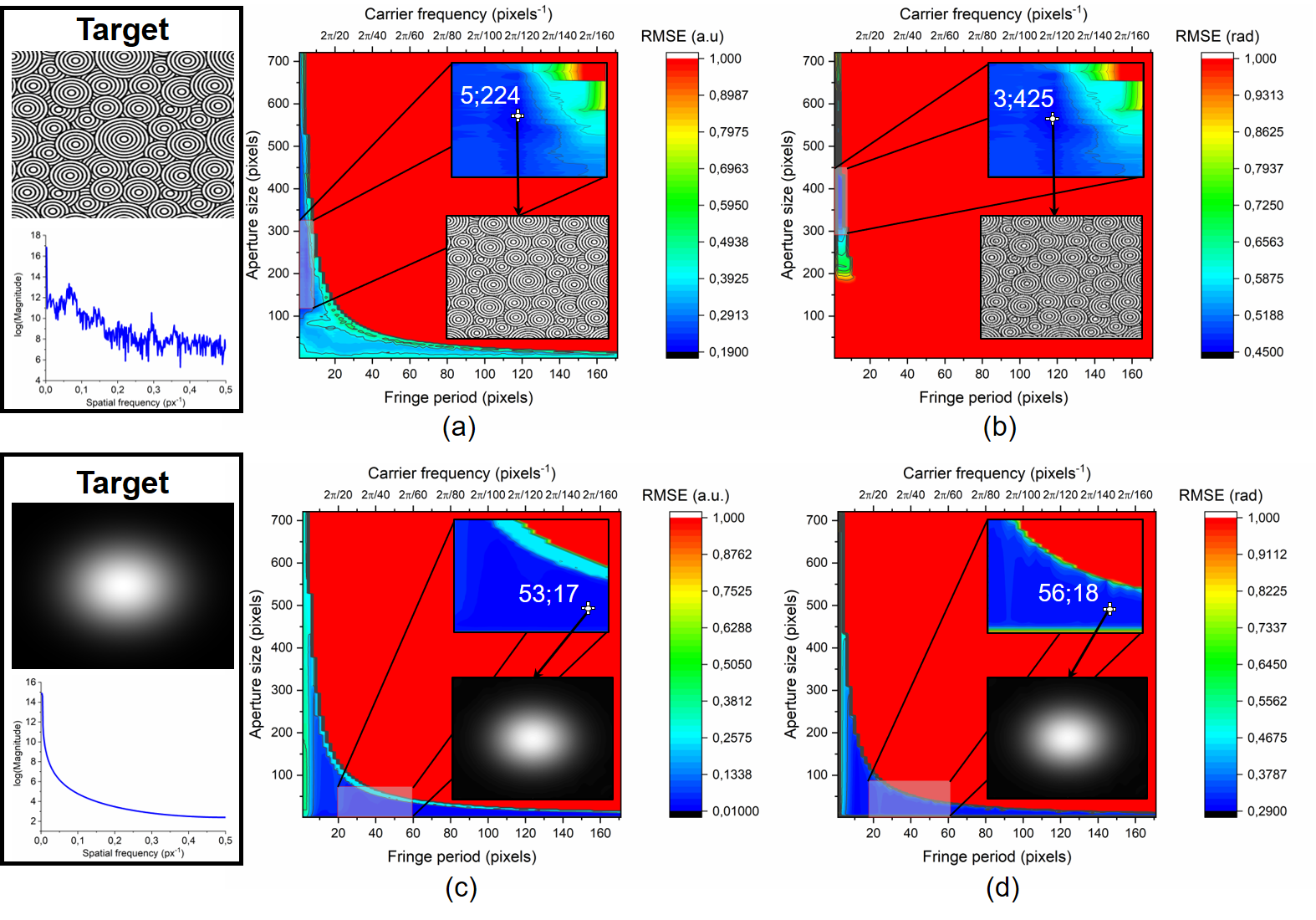}
   \end{tabular}
   \end{center} 
\caption{Amplitude (a,$\,$c) and phase (b,$\,$d) RMSE map dependent on aperture size and fringe period for the ``circles`` (a,$\,$b) and 2D Gaussian distribution (c,$\,$d) objects. Insets: enlarged fragments, obtained intensity (a,$\,$c) and phase (b,$\,$d) distributions at the points indicated by the markers. \textcolor{black}{Target distributions with corresponding angular spectrum structure diagrams are shown on the left in black rectangles.}}
%>>>> use \label inside caption to get Fig. number with \ref{}
   \label{fig:circ_gauss}
  \end{figure}

\begin{figure}[h!]
  \begin{center}
   \begin{tabular}{c} %% tabular useful for creating an array of images 
   \includegraphics[height=6cm]{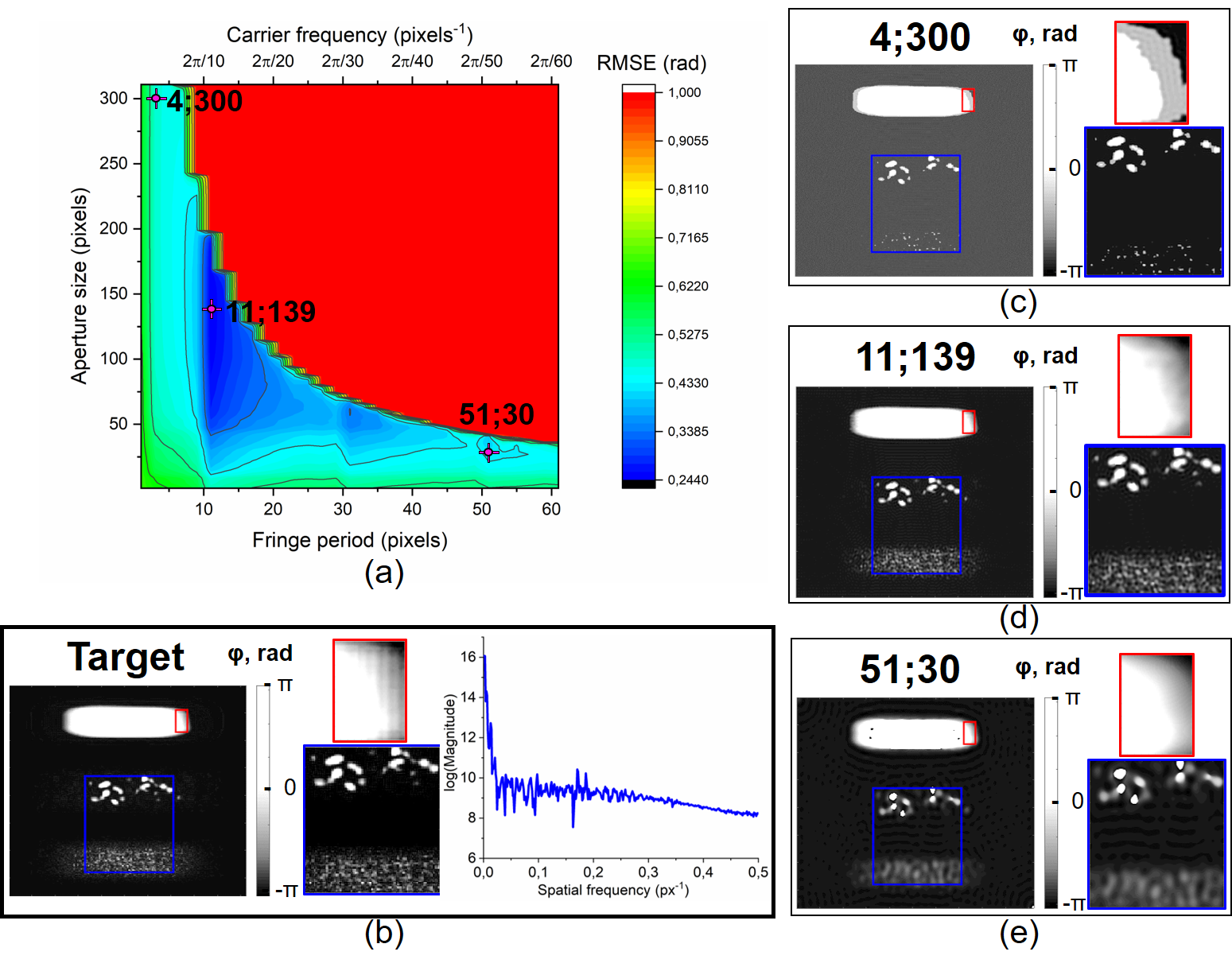}
   \end{tabular}
   \end{center} 
\caption{Phase RMSE map dependent on aperture size and fringe period for the phase object modulation (a). The RMSE map is shown on a larger scale. \textcolor{black}{(b) Target phase distribution and enlarged fragments (red and blue rectangles) with corresponding angular spectrum structure diagram are shown in a black rectangle.} Obtained phase distributions enlarged fragments obtained using the following parameters: (c) fringe period=4 pixels, aperture size=300 pixels, (d) fringe period=11 pixels, aperture size=139 pixels, (e) fringe period=51 pixels, aperture size=30 pixels.}
%>>>> use \label inside caption to get Fig. number with \ref{}
   \label{fig:falaggis}
  \end{figure}

\begin{figure}[h!]
  \begin{center}
   \begin{tabular}{c} %% tabular useful for creating an array of images 
   \includegraphics[height=6cm]{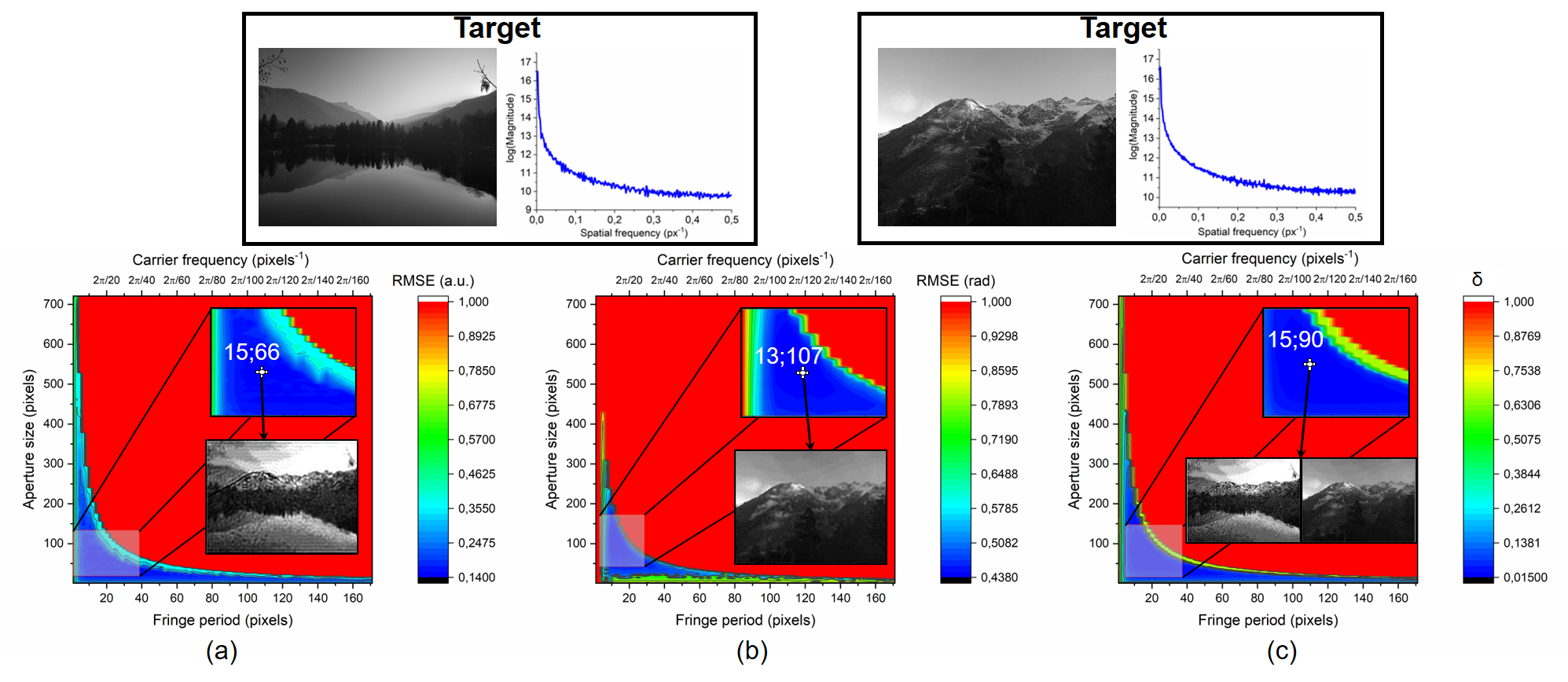}
   \end{tabular}
   \end{center} 
\caption{Amplitude (a), phase (b) RMSE and modulation error maps dependent on aperture size and fringe period for the complex wavefront modulation. Insets: enlarged fragments, obtained intensity (a), phase (b), both (c) distributions at the points indicated by the markers. \textcolor{black}{Target intensity and phase distributions with corresponding angular spectrum structure diagram are shown in a black rectangle.} }
%>>>> use \label inside caption to get Fig. number with \ref{}
   \label{fig:complex}
  \end{figure}

Let us consider the encoding of several target objects containing different range of spatial frequencies:  ``cat'', ``circles'', 2D Gaussian distribution, USAF 1951 test chart, phase object characterized by high and low spatial frequencies content (mixed-frequency phase object), which was utilized in~\cite{falaggis2015hybrid}. \textcolor{black}{The difference between objects was analyzed using radial-averaged structure of angular spectrum, which is presented in the forms of diagrams of Fourier spectrum magnitude in logarithmic scale along spatial frequency in Fig.~\ref{fig:cat_error}--\ref{fig:complex}.}
Object ``circles'' (high-frequency object) and 2D Gaussian distribution (low-frequency object) were used as the target intensity or phase distribution, while the corresponding phase or intensity distributions were two-dimensional arrays with a constant value at each point (amplitude-only and phase-only modulation). Object ``cat'' was used for amplitude-only modulation, USAF 1951 test chart, and mixed-frequency phase object were used for phase-only modulation. In addition, complex modulation was performed for ``sky'' image as amplitude and ``mountain'' image as phase target distributions.

To investigate the dependence of the modulation quality on major DMD modulation parameters, each of these images was encoded with the aperture size ranging from 1 to 720 pixels and the fringe period varying from 1 to 170 pixels. For each of the obtained images, its RMSE from the target wave was calculated and represented as a 2D pseudocolored surface (see Fig.~\ref{fig:cat_error}--\ref{fig:complex}).

In the case of ``cat'' amplitude encoding, both spatial resolution and quantization levels are crucial, as various parts of this image may have high requirements for each of these characteristics. \textcolor{black}{A smooth subsidence of the angular spectrum with the spatial frequency increment is observed. Moreover, the angular spectrum structure is characterized by both high and low spatial frequencies in the Fourier spectrum.} Therefore, the best amplitude modulation (inset 2 in Fig.~\ref{fig:cat_error}) is achieved at intermediate values of both carrier frequency and aperture size. Fig.~\ref{fig:cat_error} shows the image defects that appear in the non-optimal selection of aperture size and fringe period. Inset 1 shows insufficient amplitude levels at high spatial resolution, inset 2 shows minimal modulation error, inset 3 has low spatial resolution at a large number of grayscale levels, and inset 4 shows defects due to the passing of several adjacent diffraction orders.

The curve separating the maximum error area (orange and red colors) is hyperbola-shaped because of the inverse proportionality of the assigned fringe period and the distance between the diffraction orders. Therefore, all images obtained with parameters above this curve will be distorted by defects such as those in inset 4 due to the passing of several adjacent diffraction orders. Since various objects pose a different amount of high spatial frequencies, the hyperbola-shaped curve location varies slightly depending on the object. Thus, for further objects, only the area under this curve was considered.

The object ``circles'' is used as a demonstration of the small elements content influence on optimal aperture size and fringe period selection. The object 2D Gaussian distribution is used to demonstrate the example of high image quantization and low requirements to spatial resolution. Fig.~\ref{fig:circ_gauss} shows the maps of amplitude and phase RMSE in the case of amplitude-only and phase-only field variations, respectively, for these two objects. The insets show the examples of the obtained intensity and phase distributions at the indicated parameters, corresponding to the minimum RMSE.

In the first example, it is important to achieve the maximum possible aperture size to provide sufficient spatial resolution. \textcolor{black}{The target image includes repetitive circles of several fixed sizes. Therefore, the structure of the angular spectrum contains smooth oscillations around several values of certain spatial frequencies.}  
Simultaneously, for the second example, a decrease in the distance between diffraction orders and the aperture size does not significantly deteriorate the image quality due to the absence of small details in this image. \textcolor{black}{Its angular spectrum diagram }\textcolor{black}{is characterized by prevailing number of low spatial frequencies with almost total absence of high ones.} As shown in Fig.~\ref{fig:circ_gauss}~(a,$\,$b), the minimal error area is local and global minimum corresponds to large apertures and small fringe periods. In the case of 2D Gaussian distribution, the high error of amplitude and phase modulation corresponds only to small fringe periods (high carrier frequencies) (Fig.~\ref{fig:circ_gauss}~(c,$\,$d)). The minimum error of wavefront modulation can be achieved in case of small aperture size and large fringe period.

\textcolor{black}{Fig.~\ref{fig:falaggis} demonstrates results of phase-only modulation of an object which contain a different range of spatial frequencies, i.e., combine scattering structures with different sizes of inhomogeneities~\cite{falaggis2015hybrid,kulya2019hyperspectral}.} 

\textcolor{black}{The angular spectrum structure differs from synthesized 2D Gaussian distribution or ``cat'' object and partly resembles the structure of ``circles'' object. This can be due to the presence of small details on target distribution. Furthermore, on the graph the prevalence of low spatial frequencies is observed. Following this, there is a decline and monotonic reduction of the graph. The reason for this may be that the main part of the distribution is occupied by the band corresponding to the low frequencies elements with corresponding phase values $\pi$ and $-\pi$. Thus, these details make a significant contribution to the spectrum structure.}

The fragments containing low (red rectangle) and high (blue rectangle) spatial frequencies are demonstrated for each set of parameters. From Fig.~ \ref{fig:falaggis}~(d), it can be noted that objects with similar requirements to spatial resolution and quantization are reconstructed with minimal error in case of balanced fringe period and aperture size. Increased aperture size results in low spatial frequencies loss (Fig.~\ref{fig:falaggis}~(c)), while increased fringe period tends to loss of high ones (Fig.~\ref{fig:falaggis}~(e)).

\textcolor{black}{Objects in both amplitude and phase were selected to investigate complex field modulation. The amplitude error, phase error, and modulation error together with the resulting intensity and phase distributions are shown in Fig.~\ref{fig:complex}. The modulation error was calculated as $\delta=1-F$, where $F=|E^*_{target} E_{obtained}|^2$, $E_{target}$ is a target field, $E_{obtained}$ is the field obtained after DMD modulation\cite{goorden2014superpixel}.}

\textcolor{black}{The graphs of angular spectrum structure show that these images are characterized by the content of both high and low spatial frequencies. Therefore, in this example, it is impossible to neglect spatial resolution as well as image quantization.} The area of small errors is in the middle of the graph (insets in Fig.~\ref{fig:complex}), i.e., in the case of small apertures and small fringe periods. Increasing one or another parameter leads to an increase in modulation error.

\textcolor{black}{Due to similar types of target intensity and phase images in Fig.~\ref{fig:complex}~(a,$\,$b) global minimum of RMSE for these images corresponds to similar parameters. Meanwhile, optimal parameters of experimental setup for independent simultaneous modulation of both amplitude and phase distributions [15;93] (Fig.~\ref{fig:complex}(c)) lies in between of the values calculated for amplitude [15,66] and phase [13;107] images only.}

\textcolor{black}{It should be noted that the angular spectrum structure of the objects "sky" and "mountains" is similar to the structure of the object ``cat''. The latter was modulated as an intensity distribution, and the optimum was in the same area as for the "sky" object. In addition, regarding phase and amplitude modulation in Fig.~\ref{fig:complex}, it is shown that the complex modulation optimum is in the middle between the two coordinates for amplitude and phase modulation. Therefore, knowing the optimum parameters for amplitude and phase modulation separately is possible to predict the optimum parameters for complex modulation.}

\subsection{Experimental results}
\label{subsec:experiment}
In the presented above sections, the factors affecting the quality of amplitude or phase images are discussed. On this basis, for the experimental validation of numerical simulation, optimization of the parameters of the binary pattern and the experimental setup was conducted.

The experimental setup is described in Subsection~\ref{subsec:setup}. Experimental confirmation of the image reconstruction quality dependence on aperture size was performed on the example of phase-only modulation of the USAF 1951 test chart. This object is usually applied for the determination of optical systems spatial resolution and can be successfully used for the demonstration of aperture size impact on the spatial resolution of the modulated wave. Since this object should be reconstructed with the highest possible spatial resolution, the setup parameters and the binary pattern were selected according to the optimization results (Section~\ref{subsec:RMSE}): the aperture size diameter was 301 DMD pixels and the fringe period was 5 DMD pixels. \textcolor{black}{Considering the parameters of the experimental setup, the aperture diameter should be equal to approximately 2.27 mm.} 

To confirm the accuracy of the selected parameters, the aperture size for filtration was changed in the $x$ axis. The micrometer-adjusted monochromator slit was set in the Fourier plane to filter the first diffraction order. It allowed us to precisely vary the size of the filtration area along the $x$ coordinate. The variation of only a single coordinate spatial aperture enables the demonstration of the impact of the aperture size. 

%%%%эта пикча влезет в 1 столбик
\begin{figure}[h!]
  \begin{center}
   \begin{tabular}{c} %% tabular useful for creating an array of images 
   \includegraphics[height=13cm]{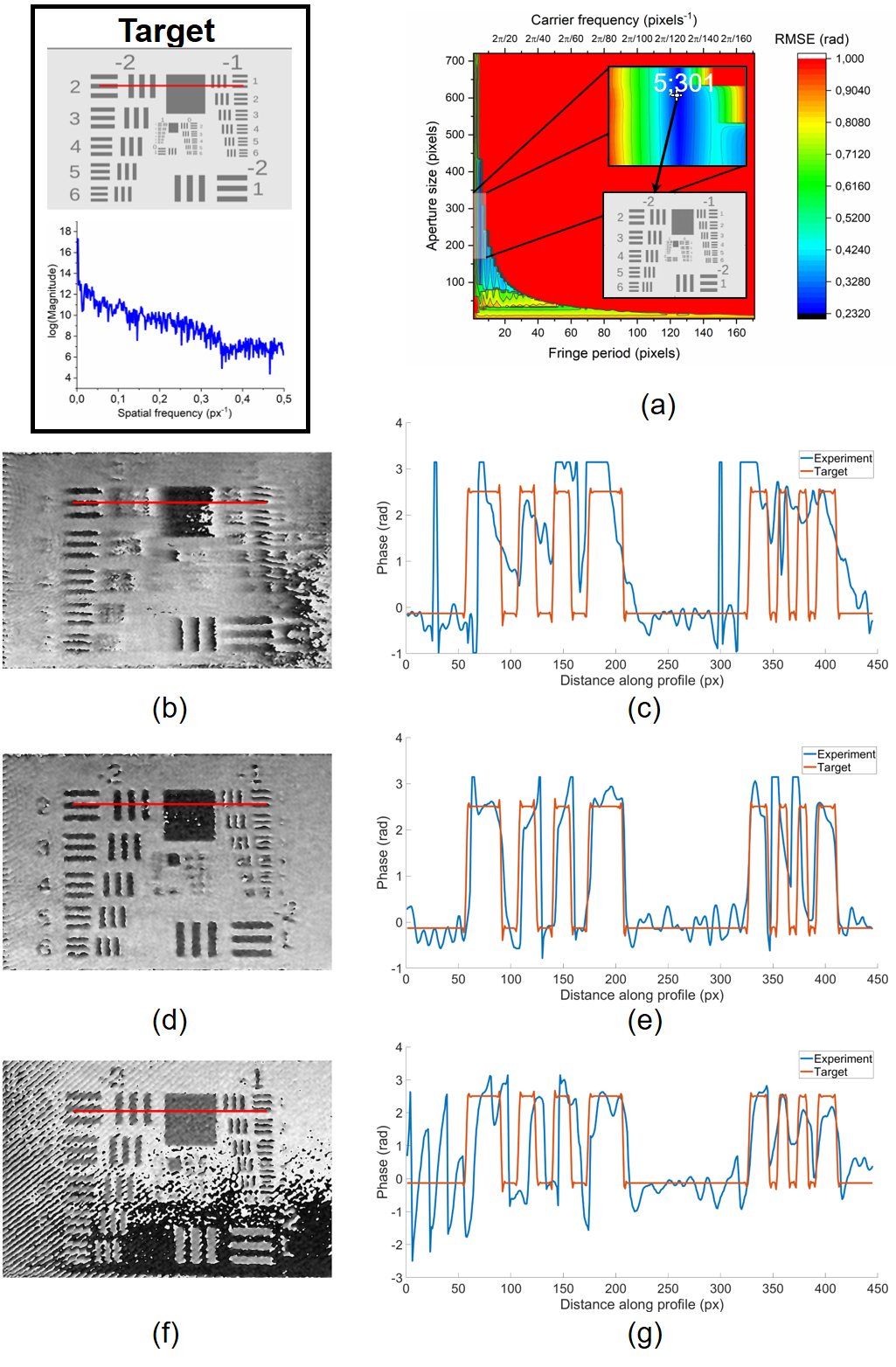}
   \end{tabular}
   \end{center} 
\caption{The results of phase distributions reconstruction with different filtering aperture sizes. (a) Phase error map for USAF 1951 test chart, (b,$\,$d,$\,$f) reconstructed phase distributions with aperture size 1.5 mm, 2.25 mm and 3.0 mm, respectively. The red lines indicate the coordinates of cross-sections; (c,$\,$e,$\,$g) the cross-sections of reconstructed phase distributions and target distribution. \textcolor{black}{Target phase distribution with corresponding angular spectrum structure diagram are shown in a black rectangle.}}
%>>>> use \label inside caption to get Fig. number with \ref{}
   \label{fig:experiment}
  \end{figure}

\textcolor{black}{It can be clearly seen on the Fourier spectrum that this distribution contains the large amount of high spatial frequencies. Consequently, it is confirmed that the maximum possible aperture size should be used.} The results of phase distributions reconstruction at different filtering aperture sizes are shown in Fig.~\ref{fig:experiment}. Whereas the size was changed along the $x$ axis, image blurring using the insufficient aperture size is observed only in the horizontal direction (Fig.~\ref{fig:experiment}~(a)). Fig.~\ref{fig:experiment}~(e) shows the image deterioration due to adjacent diffraction orders passing through the filtration aperture. As can be seen in Fig.~\ref{fig:experiment}~(c), the optimal aperture size allowed us to reconstruct the phase distribution with minimal error and high image quality. The spatial resolution was calculated using the following equation: $Resolution=2^{Group Number+\frac{Element Number-1}{6}}$ As it can be seen from Fig.~\ref{fig:experiment}~(d), Group Number was -1, and Element Number was 3. Real size USAF 1951 test chart to DMD pattern size conversion ratio was 13.8. Therefore, the estimated spatial resolution was 8.72~lp/mm. 

The cross-sections comparing the target and reconstructed distributions are provided (Fig.~\ref{fig:experiment}~(c,$\,$e,$\,$g)). In the case of the optimal aperture, an almost complete phase coincidence is observed (Fig.~\ref{fig:experiment}~(d)). For a small aperture, there are no sharp edges between the objects, which is demonstrated in Fig.~\ref{fig:experiment}~(b). For large apertures, the image corrupting fringes resulting from the adjacent diffraction orders passage can also be seen in the cross-section in Fig.~\ref{fig:experiment}~(f)). 

The experimental results have demonstrated that the parameters of the binary pattern and aperture size, selected by the described above method allowed to reconstruct the phase with high image quality. The correlation between the aperture size and image quality was also confirmed.

\section{Discussion and conclusion}

Over the past few decades, wavefront manipulation by DMD has been widely used in light processing technology, despite the fact that DMD itself can be used for binary modulation of amplitude distribution only. \textcolor{black}{The modulation of both amplitude and phase distribution is performed in two steps. Firstly, it is necessary to generate a binary pattern and project it on DMD. The pattern must be encoded in such a way that the wave reflected from DMD is divided into several diffraction orders, and the target distribution is contained in one of them. Secondly, this diffraction order is filtered by aperture, while others are blocked.} Although there are various techniques for binarization of necessary patterns~\cite{cheremkhin2019comparative,chhetri2001iterative,stuart2014fast,yang2019error,goorden2014superpixel}, computer-generated Lee holography~\cite{lee1974binary} is one of the most reliable methods to perform fast and efficient generation of binary patterns. 

As it was mentioned above, this method assumes the generation of binary patterns and filtration of the first diffraction order in the Fourier domain. The quality of the wavefront modulation is significantly affected by experimental parameters such as filtration aperture size and binary hologram carrier frequency. The variation of these two parameters enables the optimization of the quality of complex wave modulation and the achievement of either high spatial resolution or high image quantization. If the target complex wave requires high spatial resolution, carrier frequency, and filtration aperture size should be increased. However, target complex wave with gradually changing phase or amplitude values modulation requires a high quantization of target amplitude or phase distribution. Therefore, the considered parameters should be decreased to provide the best quality of wave modulation. 

The correlations between quantization and carrier frequency, resolution and aperture size are particularly interesting considering the wide variety of applications utilizing DMD. Some type of application has different requirements for spatial resolution and quantization of intensity and phase. For instance, DMD patterns with lower spatial frequencies are used to correct aberrations by increasing the fringe period and decreasing the filtering aperture. Laser processing or lithography objectives~\cite{salter2019adaptive,yoon2018emerging} require increased accuracy and high spatial resolution, which can be achieved by increasing the filtering aperture and decreasing the fringe period of binary hologram. High spatial resolution is also required in imaging applications such as optoacoustic microscopy~\cite{yu2017wavefront} or optical coherence tomography~\cite{kim2018finite}. However, as we have demonstrated in Subsection~\ref{subsec:filtration}, reduction in the binary pattern carrier frequency results in an increase of phase and amplitude quantization. In the case when the target wavefront poses a slow variation of amplitude and phase distributions, e.g., cells or intracellular organelles, quantization of modulated complex wave becomes important. Thus, in applications of such type, it is impossible to sacrifice one parameter to improve another. In this case, it is necessary to maximize equally the spatial resolution and quantization of images.

In summary, the results reported here demonstrate that the achievement of the accurate wavefront modulation using digital micromirror device requires optimization of experimental parameters according to the target complex wave and its properties related with spatial resolution of quantization levels. The mathematical and physical principles that enable independent manipulation of phase and amplitude were described and discussed in detail. Moreover, we have discussed restrictions related to the distribution of the original information capacity of the encoding binary pattern and demonstrated that trade-off between the spatial resolution of the encoded complex wave and its quantization should be achieved in dependence on the target complex wave. It is shown that increasing one of these parameters results in an inevitable decrease in the other. In addition, we have proposed an \textcolor{black}{approach} for the optimization of the generated DMD binary patterns for each amplitude and phase distribution. The \textcolor{black}{approach} is based on a simulation of the modulated complex wave generation and takes into account the requirements of the target distribution to the quantization and spatial resolution. \textcolor{black}{Preliminary estimation of the necessity to increase the aperture for filtering is performed by calculating the Fourier spectrum of the target amplitude or phase distribution, which enables the determination of the content of high and low spatial frequencies.} Furthermore, the main applications of DMD for wavefront modulation, in which it is likely to be beneficial to maximize quantization or spatial resolution, are highlighted. Being optimized with the best experimental parameters, wavefront manipulation using DMD enables the accurate independent modulation of amplitude and phase distributions, and provides outstanding opportunities in various biomedical and technological applications. \textcolor{black}{This technique is promising for implementation in various fields and applications, for instance, in the fast generation of shaped beams~\cite{mirhosseini2013rapid} or in studies of scattering media, in particular, for modulation of ultrashort laser irradiation~\cite{petrov2014investigation,belashov2019effect}.} However, the best results can be achieved only when the DMD pattern generation algorithm is optimized for certain types of amplitude and phase images, specific for a given application.

\section*{Acknowledgements}

A.B. and N.P. thank Russian Foundation for Basic Research (18-32-20215).

\bibliography{sample}

\end{document}